\documentclass[letterpaper,twocolumn,10pt]{article}
\usepackage{usenix2019_v3}

\usepackage{tikz}
\usepackage{amsmath}
\usepackage{amssymb}

\usepackage{xurl}
\usepackage{xcolor}
\usepackage{listings}
\definecolor{codegreen}{rgb}{0,0.6,0}
\definecolor{codegray}{rgb}{0.5,0.5,0.5}
\definecolor{codepurple}{rgb}{0.58,0,0.82}
\lstdefinestyle{mystyle}{
    backgroundcolor=\color{white},   
    commentstyle=\color{codegreen},
    keywordstyle=\color{magenta},
    numberstyle=\tiny\color{codegray},
    stringstyle=\color{codepurple},
    basicstyle=\ttfamily\footnotesize,
    breakatwhitespace=false,         
    breaklines=true,                 
    captionpos=b,                    
    keepspaces=true,                 
    numbers=left,                    
    numbersep=5pt,                  
    showspaces=false,                
    showstringspaces=false,
    showtabs=false,                  
    tabsize=2
}
\usepackage{svg}
\usepackage{comment}
\usepackage{multirow}
\usepackage{enumitem}
\usepackage{caption}
\usepackage{subcaption}
\usepackage{titlesec}
\titlespacing\section{0pt}{4pt plus 2pt minus 2pt}{2pt plus 2pt minus 0pt}
\titlespacing\subsection{0pt}{3pt plus 2pt minus 2pt}{2pt plus 2pt minus 0pt}

\usepackage{soul}
\setulcolor{red}

\usepackage{xspace}
\newcommand{\name}{\textsf{MGuard}\xspace}
\newcommand{\design}{\textsf{MProtect}\xspace}

\newcommand{\cpu}{interpreter\xspace}
\newcommand{\Cpu}{Interpreter\xspace}

\newcommand{\boot}{\textsf{Secure Booter}\xspace}
\newcommand{\adp}{\textsf{Binary Adapter}\xspace}
\newcommand{\tcb}{\textsf{Guardian}\xspace}
\newcommand{\secure}{sensitive\xspace}
\newcommand{\Secure}{Sensitive\xspace}
\newcommand{\shadow}{cloak\xspace}
\newcommand{\Shadow}{Cloak\xspace}

\newcommand{\exvec}{interrupt vector table\xspace}
\newcommand{\exvecs}{interrupt vector tables\xspace}
\newcommand{\shim}{\code{trampoline} segment\xspace}
\newcommand{\capy}{capability\xspace}
\newcommand{\Capy}{Capability\xspace}

\newcommand{\ptbar}{page table base address register\xspace}
\newcommand{\vbar}{vector base address register\xspace}

\newcommand{\trapped}{trapped return\xspace}
\newcommand{\Trapped}{Trapped return\xspace}

\newcommand{\paragraphb}[1]{\vspace{0.05in}\noindent{\bf #1}~}
\newcommand{\paragraphc}[1]{\vspace{0.05in}\noindent{\bf\em #1}~}
\newcommand{\code}[1]{\texttt{\small #1}}
\newcommand{\program}[1]{\textsf{\small #1}}

\begin{document}
\title{\vspace{-4em}\design: Operating System Memory Management without Access}
\date{}
\author{
Caihua Li$^{1}$, Seung-seob Lee$^{1}$, Min Hong Yun$^{2}$, and Lin Zhong$^{1}$\\
$^{1}$Yale University, $^{2}$Google
}

\maketitle

\begin{abstract}
Modern operating systems (OSes) have unfettered access to application data, assuming that applications trust them. This assumption, however, is problematic under many scenarios where either the OS provider is not trustworthy or the OS can be compromised due to its large attack surface. Our investigation began with the hypothesis that unfettered access to memory is not fundamentally necessary for the OS to perform its own job, including managing the memory. The result is a system called \design that leverages a small piece of software running at a higher privilege level than the OS. \design protects the entire user space of a process, requires only a small modification to the OS, and supports major architectures such as ARM, x86 and RISC-V. Unlike prior works that resorted to nested virtualization, which is often undesirable in mobile and embedded systems, \design mediates how the OS accesses the memory and handles exceptions. We report an implementation of \design called \name with ARMv8/Linux and evaluate its performance with both macro and microbenchmarks. We show \name has a runtime TCB 2\textasciitilde3$\times$ smaller than related systems and enjoys competitive performance while supporting legitimate OS access to the user space.
\end{abstract}

\vspace{4pt}\section{Introduction}
\label{sec:intro}

Modern operating systems (OSes) have unfettered access to application (app) data. Such access has become problematic because the OS may not be trustworthy, due to the vulnerabilities from its large attack surface~\cite{linux_threat_2021} or lack of trust in the OS provider.
In recent years, many have attempted to ameliorate this problem via constructing a secure execution environment for processes, with virtualization~\cite{chen2008overshadow, zhang2011cloudvisor, li2019hypsec, mi2020cloudvisord, li2021twinvisor,van2022blackbox}, trusted execution environment (TEE) such as ARM TrustZone~\cite{guan2017trustshadow, yun2019ndss, brasser2019sanctuary}, and isolated execution environment such as Intel SGX~\cite{baumann2015shielding, arnautov2016scone, shinde2017panoply, tsai2017graphene}.

All these developments invite an important question: \textit{can modern OSes still work without unfettered memory access}? Prior works sidestep answering it by relieving the OS from managing memory used by a protected app, e.g.,~\cite{guan2017trustshadow,van2022blackbox}, or resorting to nested virtualization, e.g.,~\cite{chen2008overshadow,li2019hypsec,mi2020cloudvisord,li2021twinvisor}. 

This paper answers this question in two parts. First, using Linux as an example, we analyze how and why an OS accesses the user space (\S\ref{sec:casestudy}). We categorize the accesses into semantic vs. non-semantic. For non-semantic accesses, e.g., demand paging, the OS does not need to ``understand'' the memory content. In these cases, the OS can do its jobs even when the memory content is encrypted. For semantic accesses, e.g., system call (syscall) argument passing, however, the OS must ``understand'' the passed arguments and therefore access the memory as cleartext. We find that all cases of semantic access are well-defined, in terms of when and where. 

Second, motivated by findings from the first part, we present a redesign of two major OS abstractions, namely virtual memory and virtual \cpu. This redesign, called \design (\S\ref{sec:design}), does not use nested virtualization or any architecture-specific hardware but relies on a small, trusted software running at a higher privilege level, called the \emph{\tcb}. Unlike prior works, the \tcb does not manage the memory used by protected apps or handle exceptions.
As a result, \design provides a cleaner and more elegant foundation for protecting user-space data from an untrusted OS.
\design's \emph{secure virtual memory} (\S\ref{sec:secure_virtual_mem}) provides an encrypted view into user space for arbitrary and non-semantic OS accesses and employs a lightweight capability system to mediate all semantic accesses. 
\design's \emph{secure virtual \cpu} (\S\ref{sec:secure_exception_handle}) protects the execution context of sensitive processes and ensures their control-flow integrity.
Along with a small set of changes to the kernel's interfaces for updating page tables and accessing user-space, \design allows the OS to manage memory without unfettered access.
Because memory management is the foundation of all OS resource management, our work suggests that a modern Unix-like OS can be trivially modified to do all its job without unfettered memory access.

To validate \design, we build a prototype based on ARMv8/Linux called \emph{\name} (\S\ref{sec:impl}).
\name consists of three components: \emph{Binary Adapter} that prepares a legacy app binary for protection, a set of small changes to the Linux kernel and its booting process, and a \tcb that runs in ARM EL3. 
\name's \tcb 
is 2\textasciitilde3$\times$ smaller than the TCB from related systems~\cite{chen2008overshadow,guan2017trustshadow,van2022blackbox}.
Our measurements with both macro and microbenchmarks show that \name achieves performance competitive with related systems
while supporting all legitimate OS access to the user space (\S\ref{sec:eval}).
\section{Case Study with Linux Kernel}
\label{sec:casestudy}

We first use the Linux kernel as a case study of why and how the OS accesses user space. The case study not only motivates the design of \design but also provides necessary context to understand the limitations of related prior work. We note that others have studied user-space access by Linux kernel in the context of protecting the kernel from TOCTTOU attacks, e.g., most recently Midas~\cite{bhattacharyya2022midas}. The focus of that literature is on kernel vulnerabilities due to malicious users, not protecting users from an untrusted kernel. 

We note that related systems such as TrustShadow~\cite{guan2017trustshadow} and BlackBox~\cite{van2022blackbox} do not distinguish between semantic and non-semantic accesses: they make all memory frames used by a \secure process/container disappear from the OS.

\subsection{Non-semantic access}
\label{sec:nonsemantic}
The vast majority of cases are concerned with the kernel moving user-space data. In these cases, the kernel does not need the semantics of the data being moved. 
We call such access non-semantic.

\paragraphb{\code{read/write} syscalls}
This pair of syscalls are extensively used for data exchange with I/O, including the filesystem. The kernel is simply responsible for copying data between memory region pointed by the \code{buf} argument and the file.

\paragraphb{Demand paging}
In the case of a page fault because the kernel loads file-backed pages on demand, the kernel calls the registered \code{fault} file operation to let the device driver prepare the frame, then updates the user page table to map the prepared physical frame to the user address space.

\paragraphb{Swapping}
When the number of free physical frames is low, the kernel may swap out some cold user pages from the physical memory to the external storage like disk and invalidate the corresponding user page table entries.
When the user process accesses a swapped out page, it triggers a page fault.
The kernel swaps in the page to the physical memory, remaps it to the user address space, and lets the process continue.

\paragraphb{Memory compression~\cite{linux_mem_compression}}
Instead of swapping user pages to the external storage, the kernel can compress them after invalidating the associated page table entries.
When next time the user process access a compressed page, the kernel decompresses the page and remaps it to the user address space.

\paragraphb{Page migration}
To reduce the latency of accessing memory in a NUMA system or to mitigate the problem of memory fragmentation, the kernel may migrate the content of a physical frame to another and update the associated user page tables accordingly.
Linux provides two syscalls, \code{move\_pages} and \code{migrate\_pages}, to move user pages among nodes.

\subsection{Semantic access}
\label{sec:semantic}
In some cases, the Linux kernel does need to understand the user-space data it accesses. We say such accesses are \emph{semantic}.
All semantic accesses share three properties.
First, they are \emph{well-defined} in spatial (where) and temporal (when) boundaries.
Second, the user process knows when and where the kernel access its address space. 
Third, the kernel reads/writes the user space through well-defined interface, namely, \code{copy\_to\_user} and \code{copy\_from\_user}. These properties are the foundations to our solution of semantic access (See \S\ref{sec:cleartext}).

\paragraphb{Syscall argument passing}
The most common cases are argument passing in syscalls. For example, the kernel must understand \code{pathname} in an \code{open} syscall to open the file. 
In these cases, the kernel accesses the user space \emph{during} the syscalls (when) and in the region defined by the arguments (where).

Most syscalls pass an argument in a single transaction. That is, the kernel copies the user data specified by the argument to the kernel space when it handles the syscall. The kernel does not monitor the user data afterwards.
Prior works leverage this and copy the argument data into a buffer managed by the TCB at call time for the kernel to access later.

The \code{futex} syscall, however, is a notable exception that does not work with the buffer-based approach. When the futex syscall is invoked, the kernel syscall handler reads the user-space word specified by the \code{futex} in order to determine if and how the in-kernel wait queue for the \code{futex} should be updated. The kernel considers reading the \code{futex} and updating its wait queue as a critical section that must be done atomically, because other threads may update the \code{futex} concurrently. 

\paragraphb{Robust \code{futex}}
When a thread terminates unexpectedly while holding a \code{futex}, other threads waiting for the \code{futex} may end up waiting forever. Robust \code{futex}~\cite{robust_futex} solves this problem with a collaboration between \program{glibc} and the kernel. \program{glibc} creates and manages a list of all \code{futex}es held by the thread. The thread uses the syscall \code{set\_robust\_list} to inform the kernel where the head of the list is. When the thread terminates, the kernel traverses this list starting from its head.

\paragraphb{\code{clone} syscall}
When a user process makes a \code{clone} syscall to create a new process or a new thread, it can set the \code{CLONE\_CHILD\_CLEARTID} flag (\code{flags}) and pass the address of the child thread identifier (\code{child\_tid}) to the kernel.
When the child exits, the kernel will clear the child thread identifier by writing to the address specified by \code{child\_tid}.
Another syscall \code{set\_tid\_address} is similar.

\paragraphb{Call stack write}
In two cases, the kernel must write to the user-space memory defined by the call stack. First, when creating a process from a binary, the kernel must prepare its call stack by placing arguments at the bottom of the stack.
Second, during signal handling, the kernel prepares the call stack (or signal stack) before handing control to a user-space signal handler.
The signal handler can either use the call stack or an alternate \emph{signal stack} (also in the user space), defined by the \code{sigaltstack} syscall when the signal handler is being established using the \code{sigaction} syscall.
In both cases, the semantic write is well-defined in time and space.
\section{Related Work}
\label{sec:related}

\begin{figure}[!t]
\centering
\includegraphics[width=0.46\textwidth]{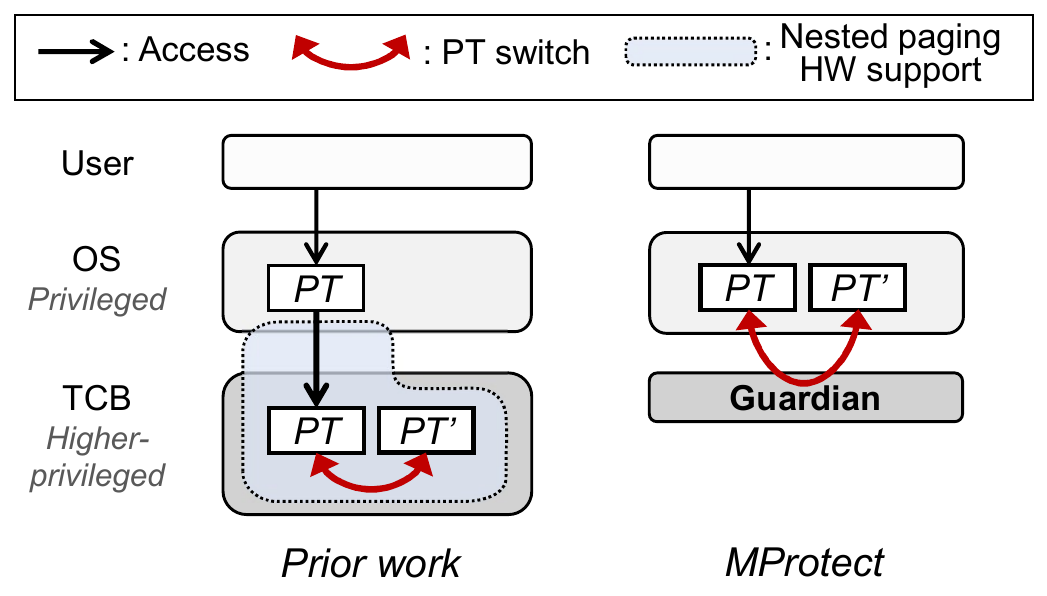}
\vspace{-2ex}
\caption{A very high-level comparison of \design's approach of mediation vs. that of indirection used in~\cite{chen2008overshadow,hofmann2013inktag,guan2017trustshadow,li2019hypsec,li2021twinvisor,van2022blackbox}. In the latter, the TCB (\textit{higher-privileged}) must manage the additional layer of address translations and its page tables (PTs). In contrast, in \design, all the PTs are still in the OS (\textit{privileged}) while the TCB (\textit{\tcb}) only mediates their use and updates, without needing special hardware support.
Similarly \design also have the interrupt vector tables in the OS while the TCB only mediates which one to use depending on whether the interrupted process is \secure or not, unlike prior work in which the interrupt vector table is inside the TCB, e.g.,~\cite{guan2017trustshadow,li2019hypsec,li2021twinvisor}.}
\vspace{-12pt}
\label{fig:design_comparison}
\end{figure}

There is an extensive literature seeking to protect app data from untrusted OSes, which motivated our investigation.
Existing works, however, took various shortcuts to achieve it without systematically addressing the OS design itself.
Such shortcuts include (\textit{i}) relieving the OS from managing the resources used by the protected app, e.g., Trustshadow~\cite{guan2017trustshadow} and BlackBox~\cite{van2022blackbox},  (\textit{ii}) relying on a large TCB below the OS, e.g., Hypervisor in Overshadow~\cite{chen2008overshadow}, (\textit{iii}) source-level application changes, e.g., Ginseng~\cite{yun2019ndss}, and (\textit{iv}) relying on hardware feature that is not widely available, e.g., ARM TZASC in Sanctuary~\cite{brasser2019sanctuary}. 
Because their goal is to protect application data, they either eschew or struggle to support legitimate access to user space by the OS.

Below we focus on three representative examples for reasons that will become clear later and refer interested readers to ~\cite{yun2019ndss,van2022blackbox} for a recent overview of this literature. 

\textit{Overshadow}~\cite{chen2008overshadow} protects unmodified applications from an unmodified OS, using a hypervisor as the TCB.
While often justly criticized for its use of hypervisor~\cite{van2022blackbox}, Overshadow pioneered two ideas. First, goal-wise, it provides the OS an encrypted view for non-semantic access to user space, a goal eschewed by much of the following work~\cite{guan2017trustshadow,van2022blackbox} but important for the OS to continue to manage resources allocated to a protected process. Second, solution-wise, it uses an additional layer of page table supported by nested virtualization hardware, which is adopted by many later works~\cite{li2019hypsec,li2021twinvisor,van2022blackbox}. As illustrated by \autoref{fig:design_comparison}, we adopt the goal but use a different solution.
Most recently, BlackBox~\cite{van2022blackbox} develops the virtualization-based solution further by substantially reducing the TCB running in the hypervisor mode and expanding the protection from a process to a container. BlackBox, however, does not allow non-semantic access by the OS: once a physical memory frame is allocated to a secure container, it disappears from the OS's view. As a result, many OS functions will stop working for these frames, e.g., swapping, memory compression~\cite{linux_mem_compression}, page migration and its related syscalls like \code{move\_pages} and \code{migrate\_pages}.

Instead of using the hypervisor mode for TCB, another line of work leverages ARM's TrustZone technology~\cite{arm2009security}. TrustShadow~\cite{guan2017trustshadow} runs a protected process inside the Secure World that is isolated from the OS in the Normal World by hardware. The memory used by the protected application is allocated in the Secure World and as a result, disappears from the OS, just like in BlackBox. 

Without systematically addressing access to user space by the OS, these works rely on ad hoc solutions that are incomplete and inefficient. 
To support syscall argument passing, they creates a per-thread buffer that is accessible by the OS and let the TCB copies data between the buffer and the user space. 
This approach is inefficient because it requires an additional memory copy.
More importantly, it requires the TCB to manage the memory used by the buffer.
Because the size of argument can be large and is impossible to know statically, the TCB has to either dynamically allocate the memory for the buffer, which increases the TCB size, or statically allocate a buffer large enough for all possible arguments, which is inefficient. 
This is particularly problematic as the buffer is also used for \code{buf} of \code{read()/write()}.

While this buffer-based approach could be used to support call-stack write, it does not work for \code{futex}. It would require that the TCB copies the \code{futex} value into the buffer, the kernel then reads the buffer and updates the wait queue atomically, which is extremely difficult if possible at all.
Not surprisingly, existing works support the \code{futex} syscall in various hackish ways. For example, Overshadow implements the futex syscall handler inside the TCB (the hypervisor) while BlackBox modifies the OS so that the OS will invoke the TCB to access the futex.

\design's goal of OS memory management without access is related to that of confidential virtual machines (VM) in HypSec~\cite{li2019hypsec} and TwinVisor~\cite{li2021twinvisor}: hypervisor memory management without access.
\design's goal, however, is more difficult than confidential VM, for two reasons. First, the interface between the OS and user processes is much wider and richer in semantics than that between a hypervisor and virtual machines. There are 55 hypercalls in Xen~\cite{xen_hypercall, xen_hypercall_list} and 27 in Hyper-V~\cite{hyperv_hypercall} while more than 400 syscalls in Linux~\cite{linux_syscall,linux_syscall_list}.
Second, like Overshadow~\cite{chen2008overshadow}, confidential VM works leverage nested virtualization support, i.e., the additional address translation for VMs, which is eschewed by \design.
\begin{table*}[!htbp]
\caption{Popular modern architectures support the architectural requirements of \name.}
\vspace{-2ex}
\label{tab:arch_req}
\centering
\small
\begin{tabular}{|c | l | l | l|} 
\hline
Requirement & x86 & ARM & RISC-V\\\hline\hline
A1 & Hypervisor mode & Hypervisor and monitor modes & Machine mode \\
A2 & VM control structure (VMCS) & TVM in \code{hcr} register & TVM in \code{mstatus} register  \\
A3 & Hypervisor call (\code{vmcall}) & Trap of control register (e.g., \code{ctr}) access & Environment call (\code{ecall}) \\
\hline
\end{tabular}\vspace{-10pt}
\end{table*}

\section{Threat Model and Design Space}
\label{sec:goal}

Before presenting \design, we first describe the threat model it defends against, the constraints we subject it to, and the requirements it demands from the underlying hardware and software system, which together define its design space. 

\paragraphb{Trusted Computing Base (TCB) and Threat Model}
We trust the hardware and software in a higher privilege level than the OS, called the \tcb. 
We also trust the tools and libraries used by the developers.
We do not trust the OS or any other software running in lower 
privilege levels.

We aim at protecting the confidentiality and integrity of application data, against any adversaries that can compromise the OS.
Toward this goal, we must also protect code integrity and control flow integrity of the applications, to fend off attacks such as return-oriented programming.
As a side effect, we protect application code confidentiality as well.

We do not protect the data sent to or received from the outside of a process, such as I/O and inter-process communication. Like the authors of HypSec~\cite{li2019hypsec} and Blackbox~\cite{van2022blackbox}, we believe such data is better protected end-to-end~\cite{saltzer1984end}.

We defend against memory mapping related Iago attacks, which means it checks the return value of system calls like \textit{mmap} and \textit{brk}.
However, \design does not defend against denial-of-service (DoS) attacks. Physical and side-channel attacks are out of scope of this paper.
Unlike solutions that keep data encrypted in memory, e.g., Ginseng \cite{yun2019ndss} and those based on Intel SGX, we do not defend against cold boot attacks as it stores data in cleartext in memory.

\paragraphb{Constraints}
\design has two absolute constraints. 
(\textit{i}) First, it must support unmodified applications. We believe that protecting application data should be transparent and orthogonal to app development and should support app binaries that already exist.
(\textit{ii}) Second, it must support  all major architectures and therefore eschew features that are available only in some specific architectures, e.g., TZASC in ARM.

Beyond the two absolute, \design must balance between changes to the OS and the size of the runtime TCB. 
While it is desirable to keep both small, when we have to choose one over the other, we choose a small runtime TCB.
\design is able to achieve 2\textasciitilde3$\times$ smaller in both size of runtime TCB and OS modifications.

Protecting application data should not come with prohibitively high overhead. 
Since non-sensitive apps in the same systems may not require protection, they ideally should pay little or no performance penalty.

\paragraphb{Requirements}
\design has three requirements for the architecture. x86, ARM, and RISC-V all meet these requirements as shown in~\autoref{tab:arch_req}.

\begin{enumerate}[topsep=0.1em,itemsep=-0.8ex, leftmargin=*, label=\textit{A\arabic*.}]
    \item A programmable privilege mode higher than what the OS is running in.
    \item A way to trap updates of the virtual memory control registers into the higher privilege mode.
    \item A mechanism with which an application can invoke the higher privileged software, bypassing the OS.
\end{enumerate}

\design has three requirements for the software system.

\begin{enumerate}[topsep=0.3em,itemsep=-0.8ex, leftmargin=*, label=\textit{S\arabic*.}]

\item The system is booted securely, which means the integrity of the \tcb and the OS is guaranteed during boot up. 
After the booting, however, the OS can be compromised.

\item The sensitive data in application binaries is not available for and tamper resistant to the OS. This can be achieved with encryption, like in BlackBox~\cite{van2022blackbox}, or secure storage, like in Overshadow~\cite{chen2008overshadow}.

\item There is a way to invoke the higher privilege software before the sensitive application can run. This can be achieved via altering the entry point in binaries, like in BlackBox~\cite{van2022blackbox}, or with a special loader, like in Overshadow~\cite{chen2008overshadow}.
\end{enumerate}
\vspace{-3pt}
In \S\ref{sec:impl}, we will present an ARMv8/Linux-based prototype that meet all these requirements.
\section{Design of \design}
\label{sec:design}

The constraints, requirements, and threat model described above collectively define the design space for a system in which the \secure process data is protected from the OS. We now provide an overview of a specific point in this design space, called \design. We will elaborate the two key abstractions of \design in \S\ref{sec:secure_virtual_mem} and \S\ref{sec:secure_exception_handle} and present a prototype implementation in \S\ref{sec:impl}.

\subsection{Key Abstractions}
Fundamentally, \design secures two abstractions supported by the OS for a \secure process: virtual memory and virtual \cpu.
\design's \emph{secure virtual memory} (\S\ref{sec:secure_virtual_mem}) provides an encrypted view into user space for arbitrary and non-semantic OS accesses and employs a lightweight capability system to mediate all semantic accesses. 
\design's \emph{secure virtual \cpu} (\S\ref{sec:secure_exception_handle}) protects execution context of sensitive processes and ensures their control-flow integrity.

\subsection{Key Insights}
\label{sec:design-insight}
\design is based on two key insights. First, all software, including the OS, addresses memory \emph{virtually} by way of the MMU and page tables. Therefore, by depriving the OS of capabilities to bypass the MMU and to arbitrarily modify the \ptbar and page tables, \design controls what virtual memory the OS sees, or \emph{blindfolds} it.
While in ordinary OSes, the OS accesses the user space via the same user page table as the process itself, \design forces the OS to use its own user page table, called \shadow page table, for non-semantic and arbitrary access.

Second, the OS relies on the \exvec to handle exceptions. By depriving the OS of capabilities to modify the \exvec, \design ensures its own version of \exvec is used when an exception interrupts the execution of a \secure process. 

Both insights rely on Architecture Requirements A1 and A2. 
Any attempt by the OS to update the virtual memory control registers, including \ptbar, will trap into the higher privilege level. \design also introduces a small change to the OS where the latter updates a page table, e.g.,  \code{set\_pte()} in Linux. The change invokes software in the higher privilege level, called \emph{\tcb}. 

Simple and small, the \tcb checks the attempts by the OS and performs a small set of sensitive operations on behalf of the latter, e.g., updating the page table.
Specifically, the \tcb makes sure that frames hosting page tables and \exvecs are never mapped as \code{writable} in any page table; and frames hosting \secure user data are never mapped in page tables other than their owner's user page table.
Unlike the TCBs from prior works, \design's \tcb does not manage memory used by protected processes or handle exceptions for them.
\autoref{fig:design_comparison} contrasts \design's approach of mediation with that of indirection taken by prior work. 

By allowing changes to the OS, \design can substantially reduce the size and complexity of the TCB (i.e., the \tcb), as compared to Overshadow~\cite{chen2008overshadow}. Leveraging Architecture Requirement A3, \design further employs user-space trampolines to invoke the TCB so that the size and complexity of OS changes are also reduced, as compared to BlackBox~\cite{van2022blackbox}.
As \design must support existing binaries, such user-space trampolines can be added as a shim, either to the binary, as in BlackBox~\cite{van2022blackbox} and our prototype, or to the loaded program, as in Overshadow~\cite{chen2008overshadow}.
\design employs two user-space trampolines: one traps into the \tcb when a \secure process is created from the binary and the other when the OS returns to a \secure process after exception handling (See \S\ref{sec:secure_exception_handle}).

\subsection{\tcb}
\label{sec:guardiandesign}

The \tcb is the software TCB running in a higher privilege mode. It is similar to a hypervisor underneath the OS, which does not interfere with the OS's normal functionality except intercepting sensitive operations.
But unlike a hypervisor, the \tcb does not manage resources or handle exceptions for a protected process; nor does it rely on nested paging hardware support, so it is much smaller in code size.
The \tcb disallows any mapping to the frames hosting Guardian's own memory.
As a result, the \tcb protects itself from the rest of the system without relying on any architectural specific hardware support, e.g., ARM TZASC, beyond the MMU.
The \tcb is re-entrant and can be concurrently invoked from multiple threads running on different cores.

The \tcb identifies a process by the base address of its page table and a thread within a \secure process by the value of the stack pointer (SP).
Notably, it does not trust the unique process identifier (PID) assigned to each process by Unix-like OSes. 

\paragraphb{ABI}
The \tcb has a narrow ABI that includes seven entry points.
(1) \code{vmcr\_trap} is invoked when a virtual memory control register is being updated.
(2) \code{exception} is invoked from the secure interrupt vector table when an exception happens to a running sensitive process, which does preparation work to secure the exception handling by the OS.
(3) Three entry points are invoked directly by the OS by code inserted to the OS: \code{set\_pts} when the OS tries to update a page table; \code{free\_pgd} when after a \secure process exits, the OS has reclaimed the memory used by its page tables; and \code{memory\_move} for semantic access by the OS.
(4) The last two are invoked from user-space trampolines: \code{process\_start} when a \secure process is being created and \code{process\_resume} when a \secure process is being resumed after exception handling.

\paragraphb{Bookkeeping}
The \tcb tracks necessary information of physical frames and sensitive processes, without relying on the data structures in the OS.

\begin{itemize}[topsep=0.3em,itemsep=0.02em, leftmargin=*]
    \item \textit{physical frame information} tells if a frame hosts sensitive data and its reference count across all page tables.
    
    \item \textit{\secure process information} include per-process (1) base addresses of the page tables; (2) address of user-space trampolines; (3) secret keys for encrypted view; (4) \capy list for semantic access; (5) execution context when preempted; 
    (6) a list of 3-tuple, i.e., (start, end, permission), which represents the range of a virtual memory area and if it is read-only or read/writable.
\end{itemize}

To remain small, the \tcb 
does not take up any responsibility of memory management. The memory used for physical frame information is fixed and allocated statically since system boots, about 8 MB in our prototype with 4 GB main memory.
The memory used for \secure processes is proportional to the number of \secure processes.
The \tcb can store it in a reserved area in the user space of the process, like in Overshadow~\cite{chen2008overshadow},
or use statically allocated memory for simplicity like in our prototype (\S\ref{sec:tcb-impl}), in which a few MB would be enough for thousands of \secure processes.
 \section{Secure Virtual Memory}
\label{sec:secure_virtual_mem}

\design's secure virtual memory has two complementary parts:
(1) an encrypted view into user space for non-semantic access by the OS and
(2) a light-weight capability system that allows semantic access into the user space by the OS.
Notably, the OS is still fully responsible for virtual and physical memory management.

\subsection{Coherent Encrypted View}
\label{sec:dual_view}

Inspired by Overshadow \cite{chen2008overshadow} but using a more efficient design, \design provides an encrypted view of the user space to the OS for all accesses beyond semantic ones. 
When control transfers from the \secure process to the OS (see \S\ref{sec:secure_exception_handle}), the \tcb intervenes and ensures that the OS uses the \shadow page table by setting the \ptbar accordingly. 
If the OS attempts to read the user space without going through the \tcb or a user-granted capability, it gets the encrypted view, supplied by the \shadow page table.
A page in the encrypted view is encrypted with a process-specific key generated by the \tcb and is coherent with its counterpart in the (cleartext) user view.
\design achieves negligible overhead for the common cases in which only the process accesses its own user space, by lazily updating the encrypted view.

We next describe how \design achieves the coherence and tracks the ownership of an encrypted page for secure data movement by the OS.

\paragraphb{Coherence}
Coherence is necessary only when a page is mapped in both the encrypted and user views, each with a physical frame.
Logically, these two physical frames should host the same data, except that one is the encrypted version of the other. 
While the encrypted view is readonly for 
the OS, the process can read and write the user view by default.

\design must maintain the coherence between the two views to deal with potential updates in the user view.
\design achieves this by marking the page-table entries of a double-mapped page \code{read-only}.
When the process writes into a double-mapped page, it will produce a permission fault and trap into the \tcb, which invalidates the corresponding page-table entry in the encrypted view and then let the write go through, as the page is no longer double-mapped but user-only.
To control the frequency of transition between double-mapped and user-only, the \tcb suspends the reading OS thread for one scheduler interval when the page transits from double-mapped to user-only.

This design has two benefits.
First, it allows unfettered concurrent reads by the process and the OS.
This contrasts the design of Overshadow~\cite{chen2008overshadow} that 
switches between the two views to serialize the reads by the OS and the process, with encryption/decryption for each switch.
Second, it avoids race conditions when the OS reads a user page that is writable by the process.
We note that legitimate OS accesses to user space usually take measures to avoid race conditions themselves. For example, for page migration, the Linux kernel unmaps a page from the user page table before moving it. As a result, \design does not need to suspend the OS thread for them.

\paragraphb{Ownership and Integrity}
Once a page is in the encrypted view, the OS can move or share it as usual, e.g., page migration.
At the same time, \design must track the ownership and protect the integrity.
\design does so by maintaining a pair of process-specific secret keys, for encryption and signature respectively.
When an encrypted page is created, the \tcb generates a signature for the page and stores the signature to the reserved area in user space, indexed by the virtual page number.
On the other hand, when a page is mapped to the user space, e.g., when the OS swaps back a page from the disk, the \tcb computes the signature of the page and compares it against that stored signature.

The cryptography operations, including both encryption and signature, are expensive and lead to the major overhead in our prototype implementation of \design (\S\ref{sec:eval}).

\subsection{Capability-based Semantic Access}
\label{sec:cleartext}

Unlike prior works where semantic access is supported in a case-by-case manner, \design supports all semantic accesses with the same \capy system, without extra copying. 
The use of \capy for access control is motivated by two insights from the case study presented in \S\ref{sec:casestudy}.
First, semantic accesses to user space by the OS are well-defined in time and space, and the user process explicitly grants the permission via the system call interface. 
Second, the OS accesses the user space via a set of narrow and stable in-kernel interfaces as in both the Linux and FreeBSD kernel~\cite{access_user_space,freebsd_copyin}, for security reasons~\cite{kernel_user-space_access}.
These interfaces are stable and date back to Linux kernel v2.2 and FreeBSD v2.2.1.

In the light-weight \capy system, a \capy to access user space in cleartext is created and checked by the \tcb. 
A \capy is  a contiguous region in the user address space specified by a 4-tuple \code{(addr, size, rw, life)} where \code{addr} is the start address of the region, \code{size} size in byte, \code{rw} a bit indicating if it is read-only or read/writable for the kernel, and \code{life} either the thread identifier, i.e., the stack pointer, which indicates the lifetime of the \capy ends with the return of the exception, or the base address of the stack which indicates a long-lived capability.
The \tcb maintains a list of capabilities for each \secure process.

\paragraphb{Lifetime of \Capy}
A user process grants the kernel \capy to access its address space when it makes system calls that pass pointer-based arguments. 
Because a \secure process's system call already traps into the \tcb (see \S\ref{sec:secure_exception_handle}), the \tcb creates the \capy by extracting information from sensitive context according to the system call semantics, detailed below.
The \tcb identifies the system call from control registers and syscall ID, e.g., exception syndrome register (ESR) and general purpose register \code{X8} in ARMv8-A.
The \tcb refers to a table, translated from the system call header file, describing mappings from the syscall ID to a set of metadata specifying the 4-tuple capability.
Upon the return of the syscall, which also traps into the \tcb (See \S\ref{sec:secure_exception_handle}), the \tcb destructs the \capy that ends with the syscall, indicated by \code{life}.
Beyond the common cases of syscall arguments, robust \code{futex} and \code{clone} require long-lived capabilities as discussed in \S\ref{sec:casestudy}. These capabilities are destructed when their owner threads terminate.

\paragraphb{\Capy checking and access}
The Unix-like OSes always access user space data with care, through narrow and stable interfaces such as \code{copy\_from\_user}~\cite{access_user_space}.
\design modifies the implementation of these interfaces so that the \tcb is invoked to check the \capy.
Once the \tcb verifies the \capy, it copies the data between the user and kernel space according to the request.
Since there is no extra copying as in the buffer-based approach used in prior work~\cite{chen2008overshadow,guan2017trustshadow,van2022blackbox}, \design's approach is both more efficient and more general, supporting all cases of semantic access, including \code{futex}.
\section{Secure Virtual \Cpu}
\label{sec:secure_exception_handle}

In addition to virtual memory, a modern OS provides the abstraction of virtual \cpu~\cite{saltzer2009principles}:
a process believes its sequence of instructions are interpreted in order without disruption, while the OS can interrupt or even suspend its execution.
The mechanism used by modern OSes to interrupt the process execution is exception.
When an exception happens, the process execution is suspended.
The CPU switches into the kernel mode and the OS saves the process context to the memory (stack) and executes the exception handler indexed in the interrupt vector table by the exception type.
When the OS finishes handling the exception, it may return to a suspended process by restoring the latter's context.
This, however, gives the OS an opportunity to access the execution context or even the entire user space \emph{without} using the user page table but via compromising the control-flow integrity of the process, e.g., return-oriented programming (ROP) attacks~\cite{roemer2012return}.

\paragraphb{Suspension with Secure Interrupt Vector Table}
To secure the execution context, the \tcb must be invoked before the OS gains control.
\design realizes this by using a \emph{secure \exvec} for \secure processes by setting the vector base address register properly, in place of the \exvec used by the OS for non-\secure processes.
The secure \exvec is an exact copy of the \exvec with one extra instruction invoking the \tcb (\code{exception()} ABI) before calling the corresponding exception handler. 
The \tcb saves the context to a reserved area in the user space, which is already protected by the secure virtual memory, or to its own memory like in our prototype.
It then clears the registers, sets the \ptbar to the \shadow page table, sets the \vbar back to the original \exvec, and finally handles the control to the exception handler.

\design's approach to invoke the TCB at exception contrasts those taken by prior works. 
For example, BlackBox~\cite{van2022blackbox} uses a single, modified \exvec that traps all exceptions into the TCB regardless of whether the running process is \secure or not. 
\design's design of dual vector tables avoid unnecessary overhead for non-sensitive processes. 
Ginseng~\cite{yun2019ndss} modifies the \exvec at \emph{runtime} to invoke the TCB, incurring higher overhead at runtime.

\paragraphb{Resumption with Trapped Return}
When the OS resumes the execution of a \secure process after it handles an exception, it simply returns control to the process starting with the saved program counter.
There is no obvious point that the \tcb could intervene.
\design solves this problem with a simple mechanism called \emph{\trapped}. 
When the \tcb clears the context after saving it to the secure stack, it sets the exception link register, which is supposed to contain the program counter of the interrupted process, to the user-space trampoline that invokes the \tcb ABI \code{process\_resume()}.
As a result, when the OS returns control back to the process, it unknowingly returns control to the code that invokes the \tcb.
The \tcb will then restore the actual context from the secure stack, switch the cloak page table back to the user page table and the normal \exvec back to the secure one, before returning control to the process resuming from the saved program counter.

\Trapped does not require OS modification and is more efficient, flexible, and general than the solutions adopted by prior works. For example, BlackBox~\cite{van2022blackbox} traps all returns from exception handling into the TCB, regardless of whether the process is \secure or not. \design uses \trapped only for \secure processes.
\section{Implementation of \name}
\label{sec:impl}

In order to validate \design, we prototype it based on Linux kernel v5.5 and ARMv8-A architecture.
The prototype, called \name, is a refinement of \design because it must make 
choices that are left open by \design.
\name has a smaller runtime TCB and fewer OS modifications, i.e., 2\textasciitilde3$\times$ smaller number of lines of code (LOC), compared to related works~\cite{chen2008overshadow, guan2017trustshadow, van2022blackbox} that also support legacy binaries.

\begin{figure}[!t]
\centering
\includegraphics[width=0.476\textwidth]{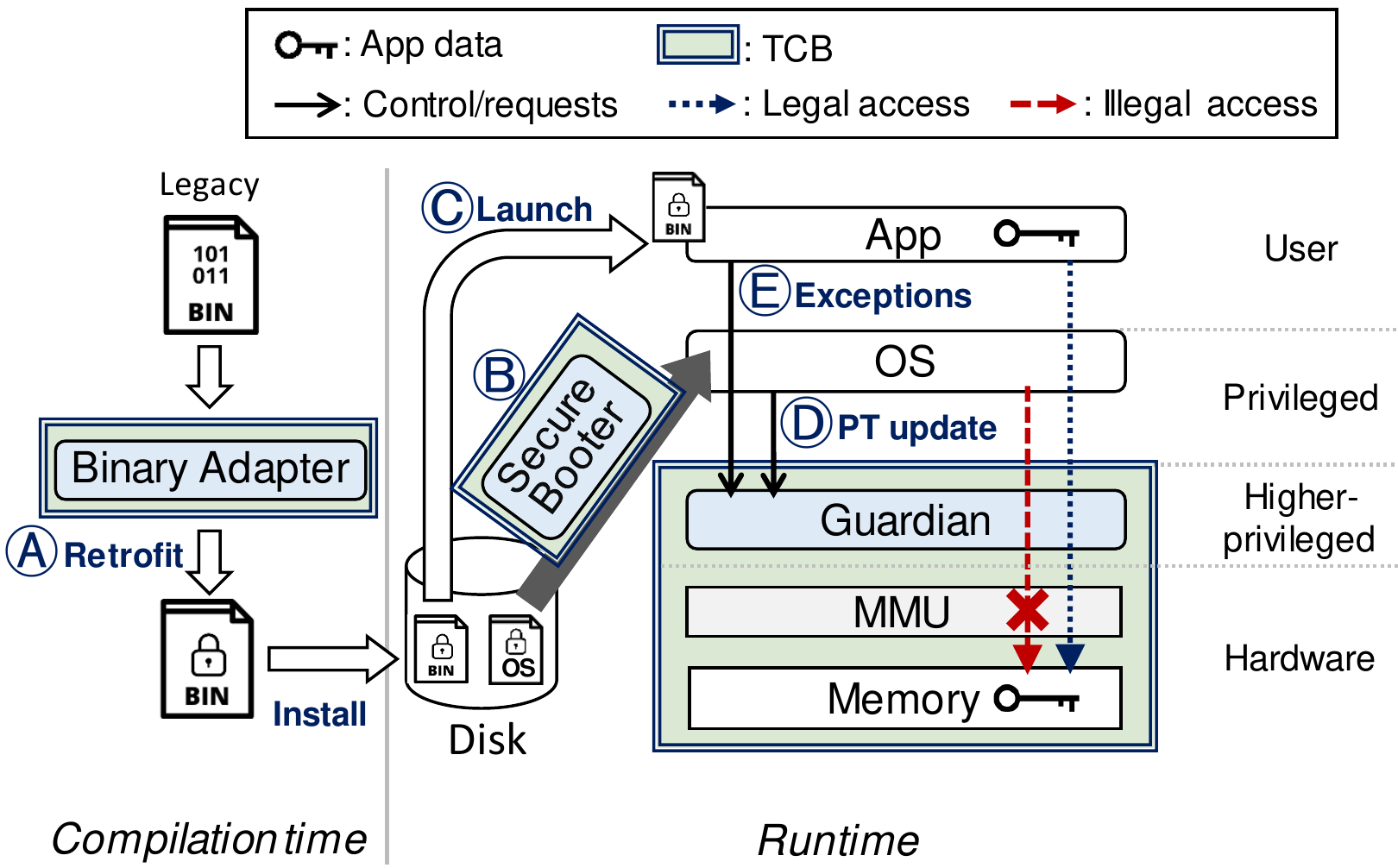}
\vspace{-20pt}\caption{\textbf{Overview of \name.} \name's \adp retrofits and encrypts binary for a sensitive application. The \tcb protects data from OS's access with mediation. (PT: page table, MMU: memory management unit)}
\vspace{-10pt}
\label{fig:overview}
\end{figure}

\name consists of three parts: \adp, \boot and \tcb as shown in \autoref{fig:overview}. 
The \adp prepares the binary by encrypting all loadable segments and adding a segment for the two user-space trampolines (\S\ref{sec:design-insight}) as well as other segments to be used by the \tcb (\textcircled{A}).
The \boot prepares the entire system via secure boot so that the OS cannot bypass the MMU or arbitrarily switch page tables (\textcircled{B}). This is achieved by configuring the processor such that the MMU is enabled and that any update to virtual memory control registers, i.e., the system control register and the \ptbar, is trapped into the \tcb (\S\ref{sec:design-insight}).
We implement \textit{\boot} as part of the ARM Trusted Board Boot process~\cite{arm_trusted_boot}, introducing 25 lines of C and assembly code to Linux kernel and 57 lines of Rust code to ARM trusted firmware.

\subsection{Binary Adaptation}
\label{sec:adaptation}
\name assumes the developer and the \tcb have their own public-private key pairs and they know each other's public key.
The developer uses the public key to prepare the binary via the \adp. The \adp generates two symmetric keys and uses them to encrypt and sign the loadable segments.
It further encrypts the symmetric keys with the \tcb's public key and embeds the encrypted keys into the binary so that the \tcb can exclusively obtain the keys to verify the integrity of the segments and decrypt them to create a \secure process.

The \adp adds the following segments into the binary and modifies the entry point in the binary header.
\begin{itemize}[topsep=0.1em,itemsep=-0.5ex, leftmargin=*]

\item The \code{signature} segment contains signatures of the loadable segments.
The \tcb uses these signatures to check the integrity of the corresponding segments.

\item The \shim includes two trampolines that invoke the \tcb ABI \code{process\_start} and \code{process\_resume}, respectively.
The \adp sets the entry point to the trampoline to \code{process\_start} so that the \tcb intervenes to create a \secure process.
The \code{process\_resume} trampoline is for resuming execution of a \secure process after returning from an exception (\S\ref{sec:secure_exception_handle}).

\item The \code{metadata} segment contains metadata used by the \tcb during sensitive process initialization.
It includes a copy of program headers, original entry point, encrypted symmetric keys, and the signature of the above data, signed with the \adp's private key.

\item The \adp further reserves a memory areas as a \code{bss} segment to store signatures generated at runtime.
\end{itemize}

We note that our \shim only contains four instructions, which is far smaller than the shim introduced in Overshadow\cite{chen2008overshadow} by 3 to 4 orders of magnitude.
Our implementation of \textit{\adp} includes 256 lines of C code, plus a cryptography library with 2.3K lines of C and assembly code imported from the Google BoringSSL project \cite{boringssl}.

\subsection{OS Modification}
\label{sec:impl_os}

To realize \design's design, \name introduces 150 LOC to the Linux kernel, which is half of that in TrustShadow~\cite{guan2017trustshadow} and one third of that in BlackBox~\cite{van2022blackbox}. 
Most modifications are for the secure virtual memory (\S\ref{sec:secure_virtual_mem}) and take place below the interfaces and data structures that have been stable since early days of the kernel.
We next describe the few key changes.

\paragraphb{Trapping Page Table Updates~(\S\ref{sec:design-insight})}
\design requires that all page table updates must be trapped into the \tcb. To achieve this, \name modifies the implementation of low level stable kernel interfaces like \code{set\_pte} to add trampolines into the \tcb.
We note these interfaces have been stable since Linux kernel 2.0 and the modification is 40 LOC.
Because Linux pre-fetches up to 16 pages around the address that triggers a read data abort~\cite{linux_mmap_few_pages},
instead of trapping page table updates one by one in \code{set\_pte}, we add another two low level interfaces, \code{set\_ptes} that batches page table updates into a per CPU buffer and \code{commit\_updates} that invokes the \tcb to serve the batched requests and empties the buffer, to be compatible with the kernel optimization.
We note that this optimization only works when multiple page table updates are made in the same critical section.

\paragraphb{\Shadow Page Table for Encrypted View~(\S\ref{sec:dual_view})}
\design mandates that all OS attempts to read the user space, unless going through the \tcb with a user-granted capability, must get the encrypted view, therefore requiring the use of the \shadow page table.
To achieve this, \name modifies the Linux kernel at three places with a total of 70 LOC.
First, we add the base address of the \shadow page table, namely \code{c\_pgd}, along side that of the user page table, i.e., \code{pgd}, in the memory descriptor \code{mm\_struct}.
By default, the \code{c\_pgd} of the \textit{init process} is set to zero and a forked process inherits the \code{c\_pgd} value from its parent in \code{dup\_mm} during process creation.
As a result, a process by default is non-sensitive and its \code{c\_pgd} is zero.
When preparing a \secure process, the \tcb sets its \code{c\_pgd} to a non-zero magic number.
Second, we modify the kernel logic to create a \shadow page table for a \secure process (non-zero \code{c\_pgd}), when it is either loaded from a \secure binary or forked from a \secure process.
Finally, we modify the kernel logic for user-space page fault handling: the kernel updates the \shadow page table if the page fault is triggered by the OS, which is indicated by a bit in the \code{flags} parameter of the page fault handler, i.e., \code{FAULT\_FLAG\_USER}.

\paragraphb{Semantic Access~(\S\ref{sec:cleartext})}
To support semantic access by the OS to the user space, we modify the implementation of intra-kernel interfaces like \code{copy\_from\_user}~\cite{access_user_space}, which have been stable since Linux kernel v2.2. The added code invokes the \tcb to check the \capy and perform the copy on behalf of the kernel.
This incurs a modification of 12 LOC.

\subsection{\tcb Implementation}
\label{sec:tcb-impl}

We implement the \tcb on top of the ARM trusted firmware that  runs in EL3.
Unlike related works also based on ARM~\cite{li2021twinvisor,brasser2019sanctuary,guan2017trustshadow}, our implementation does not rely on any ARM-specific feature, e.g., TZASC, or any software in the Secure World beyond the ARM trusted firmware. 

Our implementation clearly separates knowledge from logic and safe Rust from unsafe code. Knowledge includes 121 lines of safe Rust code to generate capabilities for system call argument passing.
Logic-wise, the implementation includes 1200 lines of safe Rust, 340 lines of assembly code / unsafe Rust for privilege level switching and direct access to hardware registers.
The \tcb also uses the cryptography library imported from the Google BoringSSL project~\cite{boringssl}.

Because related systems like BlackBox support different sets of features, it is only fair to compare the size of TCB code that supports the common features: our mediation of page table and the capability system are 350 and 270 LOC respectively, while their counterparts in BlackBox~\cite{van2022blackbox}, i.e., NPT management and syscall interposition, are both 1,000 LOC.
Moreover, \design requires only seven TCB ABIs, less than half of BlackBox~\cite{van2022blackbox} for the reason we will reveal later.

\paragraphb{\tcb ABI Invocation}
As described in \S\ref{sec:guardiandesign}, \tcb's ABI can be invoked by hardware, the kernel, and the user space.
When \code{TVM} in \code{HCR\_EL2} is set, any attempt to modify virtual memory control registers will be trapped into the \tcb (\code{vmcr\_trap}) with EL2 as a stepping stone (A2 in \S\ref{sec:goal}).
The secure \exvec and the kernel invokes the \tcb ABI (\code{exception}, \code{set\_pts}, \code{free\_pgd}, and \code{memory\_move}) with the \code{SMC} instruction.

There is no instruction on ARMv8-A that a \secure process (in EL0) can use to directly invoke the \tcb ABI. 
As a result, we use EL2 as a stepping stone for invoking \code{process\_start} and EL1 as an optimized version for \code{process\_resume} (A3 in \S\ref{sec:goal}).
On our silicon, EL1 is much more efficient than EL2 as the stepping stone, as is obvious from the measurement presented in \autoref{tab:mode_switch}.
To leverage this, we implement the \exvec switch slightly different from the design of secure interpreter (\S\ref{sec:secure_exception_handle}).
The switch takes place in \code{vmcr\_trap} depending on if the associated process is sensitive, instead of switching in every suspension and resumption.
And thus, since the secure \exvec is in use during trapped return, a \secure process can be trapped into the \tcb (in EL3) via invoking a non-existing system call to trigger an exception.
This optimization does not work for invoking \code{process\_start} when a \secure process is being created, because the secure \exvec is not yet in use.
When the \code{TID2} in \code{HCR\_EL2} is set, a \secure process can be trapped into EL2 by accessing \code{CTR\_EL0}, which is redirected to the \tcb with a \code{SMC} instruction in the EL2 \exvec.

We note that Ginseng~\cite{yun2019ndss} maps a user-space address to a frame in TrustZone's Secure Memory; the \secure process accesses this address to be trapped into EL3. Secure Memory is ARM-specific and we have found the trapping does not work on our silicon Hikey 960 because TZASC or DDR controller details are undocumented~\cite{hikey960_secure_memory_problem}.
On the other hand, \design requires less ABIs provided by the \tcb, i.e., less than half compared with BlackBox~\cite{van2022blackbox}.
It is because the \tcb collects information when it intervenes system calls, instead of exposing more ABIs for OS to invoke and provide such information.
For example, the \tcb collects memory mappings when it intervenes system calls like \code{mmap} and \code{brk}, while BlackBox~\cite{van2022blackbox} introduces an extra ABI \code{set\_vma} for the OS, which potentially adds more OS modifications.

\paragraphb{\Secure Process Creation}
When the OS transfers control to the entry point of a process as the last step of process creation, it unknowingly invokes the \tcb \code{process\_start} ABI because the entry point has been set to the user-space trampoline. The \tcb hides frames hosting the \code{metadata} segment from the OS, i.e., invalidating the corresponding kernel page table entries.
It then checks the integrity of the data with the \adp's public key and extracts the symmetric keys with its own private key.
With the metadata, the \tcb parses the program headers and walks the list of virtual memory area descriptors in the kernel space to learn the virtual addresses of other segments. It then walks the user page table. If a page is present, the \tcb hides the mapped frame from the OS, checks its integrity and decrypts its content with the extracted keys.
Finally, the \tcb configures the \vbar (\code{VBAR}) to the address of the secure \exvec (see \S\ref{sec:secure_exception_handle}) and assigns a magic number to \code{c\_pgd} before returning control to the original entry point. From then on, the sensitive process is under runtime protection from the \tcb.

\paragraphb{Page faults in \tcb}
Because \design lets the OS handle page faults, the \tcb must either avoid them or trick the OS to handle them. 
First, \name loads the entire image of the \tcb in secure boot and disables swapping for these memory so that it stays in the physical memory after booting.
This keeps the \tcb's logic simple and keeps it small.
Second, the \tcb may trigger a page fault when it accesses the user space of a \secure process per the request from the kernel, e.g., semantic access. In this case, the \tcb delegates the page fault to the OS and applies trapped return to it, via setting the exception link register (ELR) to where the OS makes the request. So the OS will re-request after the handling without extra OS modifications.
\section{Evaluation}
\label{sec:eval}

With \name as a prototype of \design, 
we report \name's performance with both macro and micro benchmarks.
The evaluation answers the following questions about \design and \name.
\begin{itemize}[topsep=0.1em,itemsep=-0.4ex, leftmargin=*]
    \item Can \name manage memory without access?
    \item What is the end-to-end overhead for both protected and unprotected processes?
    \item How much overhead does each design component contribute?
\end{itemize}

We evaluate \name with Hikey 960 development platform, which is one of the few platforms supported by the ARM trusted firmware~\cite{hikey960_arm_tf}. 
It features the big.LITTLE architecture with 4 Cortex A73 + 4 Cortex A53 and 4~GB DRAM.
We choose Hikey 960 for its combination of low cost, high computational power, and documentation availability. 
We isolate the LITTLE cores from the general scheduler with kernel boot parameter \code{isolcpus}, and pin our benchmarks to them with \code{taskset}.
We configure these cores to run at their maximum frequency of 1.844 GHz.
This setup allows our measurements to be consistent.

We focus on the runtime overhead because the overhead of \boot (\S\ref{sec:impl}) and \adp (\S\ref{sec:adaptation}) is one-time and small: our \boot adds only 40~ms to the booting process and the binary adaptation takes less than 0.1~s for programs used in our evaluation.

\subsection{Kernel functions with memory access}
\label{sec:eval_coverage}
We first show that the Linux kernel can fulfill its functions under the restrictions imposed by \name. Not all of these functions are supported in related systems.

\paragraphb{Non-semantic access}
We evaluate the non-semantic access with a system call \code{migrate\_pages}.
As described in \S\ref{sec:casestudy}, \code{migrate\_pages} triggers the OS to migrate all pages of a process from one node to another.
Because the \code{migrate\_pages} handler avoids moving pages inside the same node, we have to force the kernel to copy pages from one set of frames to another.
In more than 1000 page migration, all the movement succeed and the sensitive process continues running correctly.
We note that related systems TrustShadow~\cite{guan2017trustshadow} and BlackBox~\cite{van2022blackbox} do not support non-semantic access because memory allocated to \secure processes/containers disappear from the OS view.

\paragraphb{Supporting legacy binaries}
\name supports legacy binaries. In order to understand \name's overhead, we extensively test \name with binaries for a few microbenchmarks and four applications: \program{Nginx}, \program{Redis}, one-time password generator~(\program{OTP}), and deep neural network-based object classification~(\program{DNN}). 
\program{Nginx}~\cite{nginx} is a web server and \program{Redis}~\cite{redis} an in-memory key-value database.
\program{OTP} is based on the open source code of Ginseng \cite{yun2019ndss}. As part of two-factor authentication, it generates a one-time password by performing HMAC-SHA1 on current time and a secret key. \program{DNN} is based on the example code for DNN-based object classification from OpenCV \cite{opencv-DNN-example} and the pre-trained GoogLeNet model \cite{googlenet}.
These four applications use kernel functions in various ways:
\program{OTP} and \program{DNN} create a new process for each request;
\program{DNN} and \program{Redis} are memory intensive while Nginx is I/O intensive.
Both Redis and Nginx heavily use syscalls for I/O and timing operations.
\subsection{Understanding the overhead}
\label{sec:eval-overhead}
Like all systems intended for protecting application data, \name's protection does not come free. We next seek to understand the overhead imposed by \name using the four applications described above and necessary microbenchmarks. 

\begin{figure}[t]
\hspace*{-0.15in}
\centering
\hfill
\begin{subfigure}{0.15\textwidth}
\includegraphics[width=\textwidth]{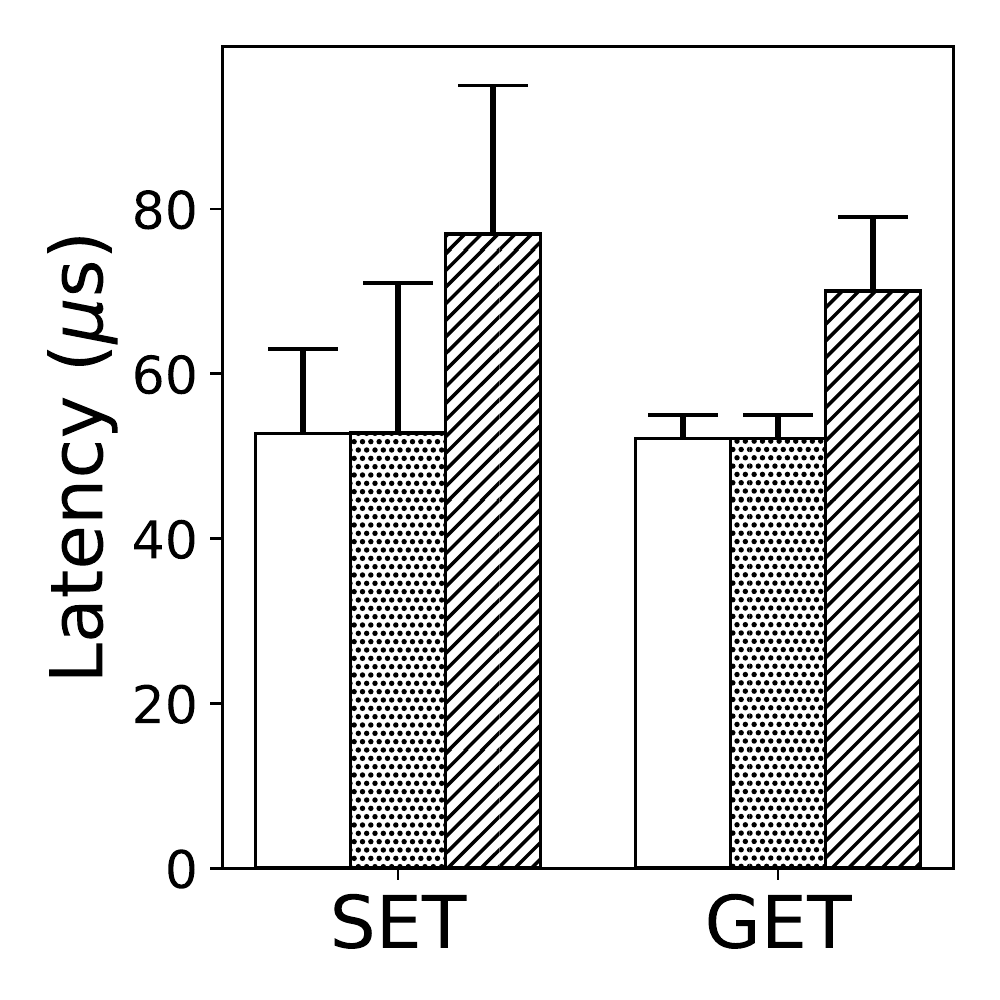}
\caption{\program{Redis}}
\label{fig:eval_redis}
\end{subfigure}
\hfill
\begin{subfigure}{0.15\textwidth}
\includegraphics[width=\textwidth]{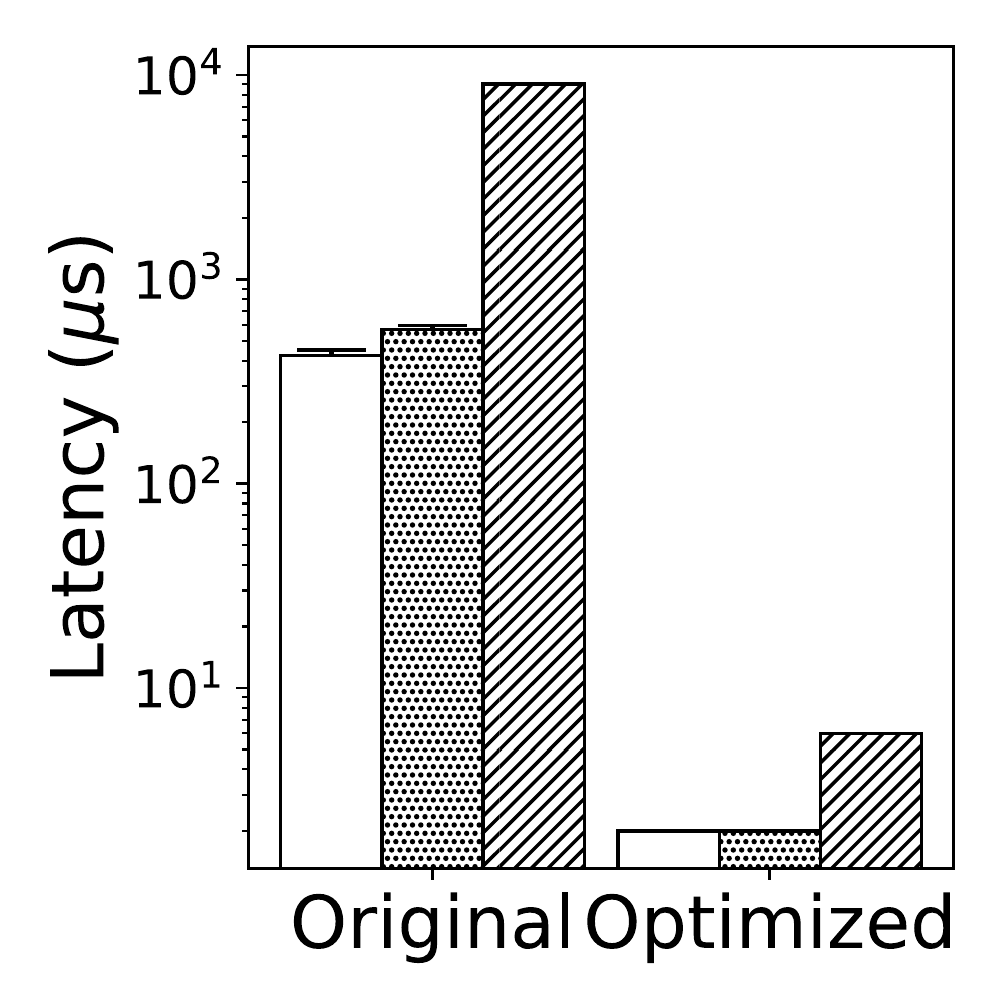}
\caption{\program{OTP}}
\label{fig:eval_otp}
\end{subfigure}
\hfill
\begin{subfigure}{0.15\textwidth}
\includegraphics[width=\textwidth]{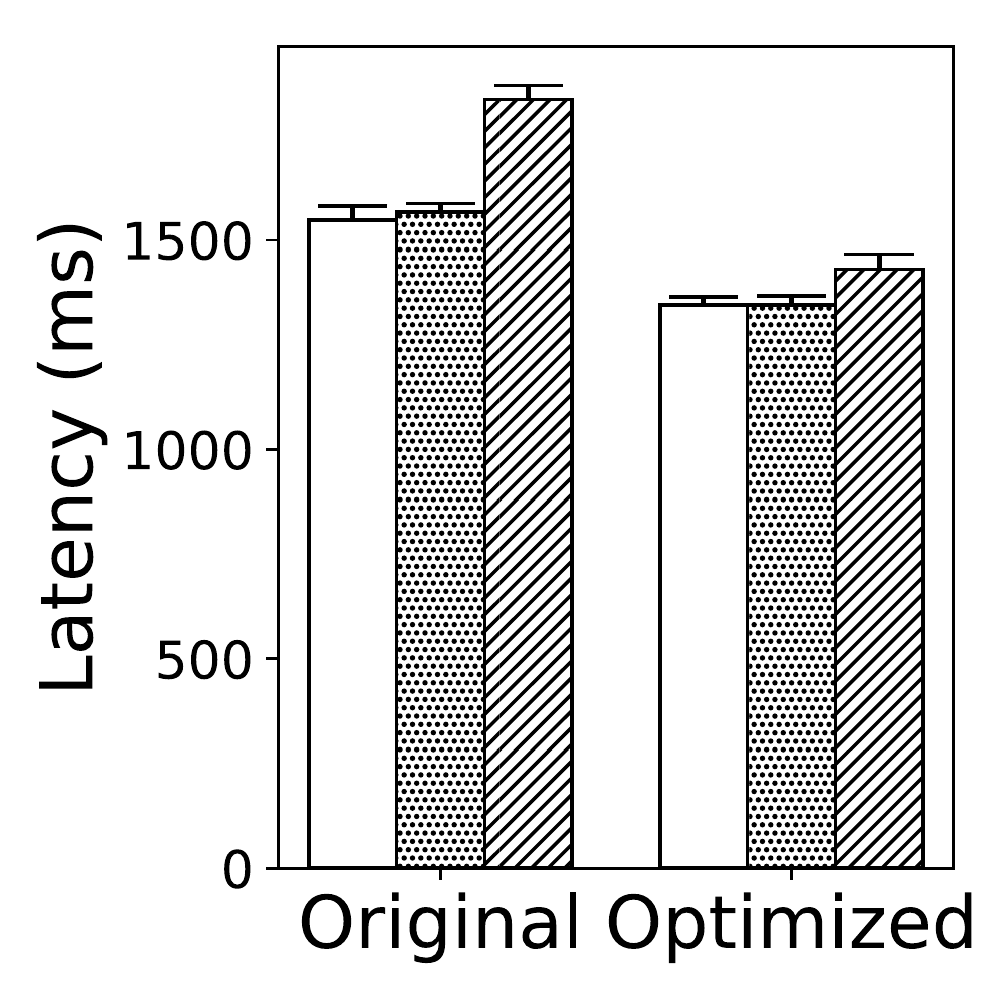}
\caption{\program{DNN}}
\label{fig:eval_dnn}
\end{subfigure}
\vspace{-10pt}
\caption{Latency. The three bars represent Vanilla, Non-sensitive and Sensitive respectively.}\vspace{-10pt}
\label{fig:eval_overhead}
\end{figure}

In order to understand the overhead of \name on both \secure and non-sensitive processes, we report results for three cases: vanilla Linux (\program{Vanilla}), non-sensitive in \name (\program{Non-sensitive}), and sensitive in \name (\program{Sensitive}). 
Whenever possible, we include the average and 99th percentile over 1000 measurements. \autoref{fig:eval_overhead} shows the latency data for all four applications. 

\paragraphc{Negligible runtime overhead for unprotected applications}:
By comparing the results for \code{Vanilla} and those for \code{Non-sensitive} in \autoref{fig:eval_overhead}, it is clear that \name imposes negligible overhead for non-sensitive processes.
That is, applications that choose not to use \name's protection pay little performance tax.
It is important because not all applications distrust the kernel.
This contrasts with the non-negligible overhead of running unprotected containers in BlackBox~\cite{van2022blackbox}, especially for applications that heavily rely on system call services, e.g., Nginx.

\begin{figure}[t]
\centering
\includegraphics[width=0.4\textwidth]{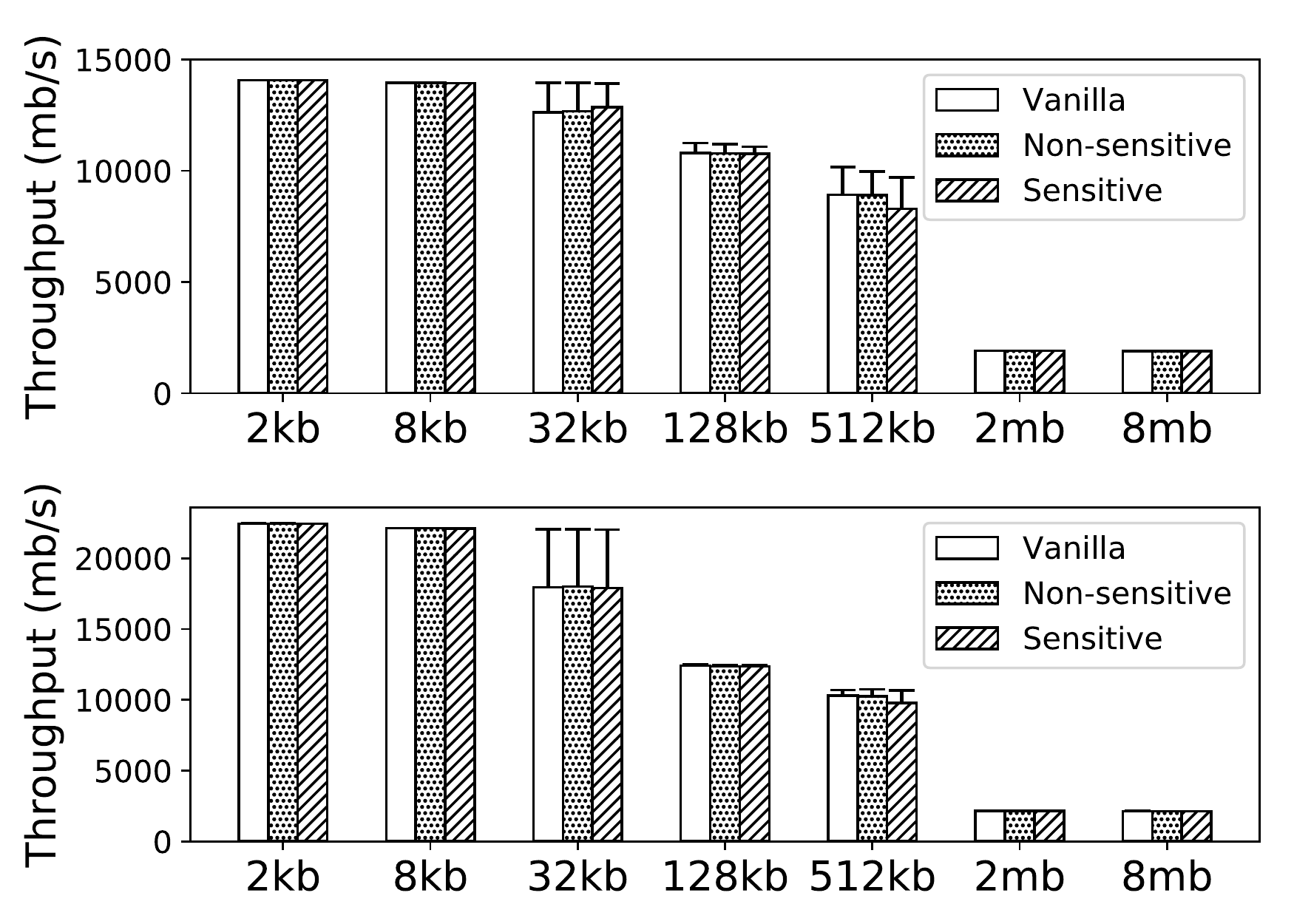}
\vspace{-10pt}
\caption{Memory throughput: read (top) and write (bottom).}
\label{fig:mem_access}
\vspace{-15pt}
\end{figure}

\paragraphc{Negligible overhead for application's own memory access}:
\design's secure virtual memory design aims to impose as little overhead as possible for the common case in which a \secure process accesses its own memory. 
The latency data for memory-intensive \program{DNN} already suggest such overhead is small. We further single out the performance of memory access using \program{bw\_mem} from LMbench~\cite{lmbench, mcvoy1996lmbench}.  \program{bw\_mem} measures the throughput of memory read and write. As confirmed by \autoref{fig:mem_access}, the overhead of \name is negligibly small.

\paragraphc{Sources of overhead for \secure processes}:
Using more microbenchmarks from LMbench, we are able to analyze the major sources of overhead for \secure processes.

\begin{figure}[!t]
\centering
\hspace*{-0.2in}
\includegraphics[width=0.35\textwidth]{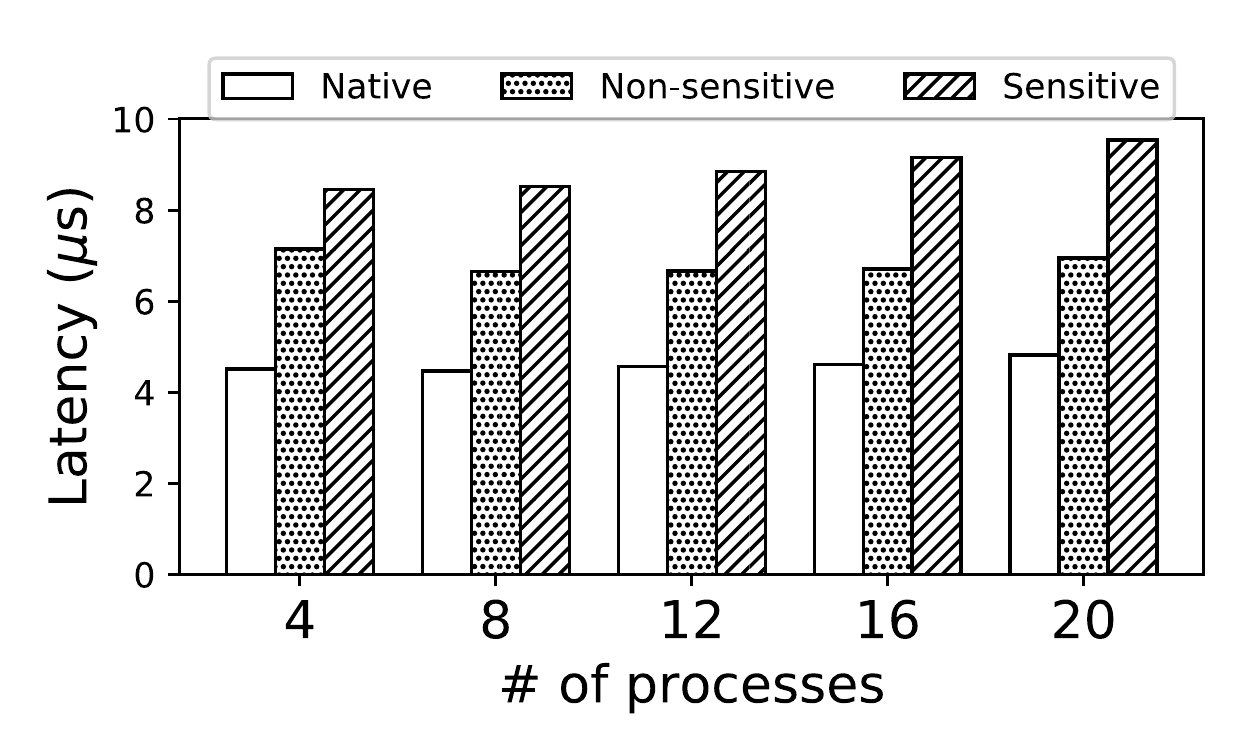}
\vspace{-12pt}
\caption{{Latency of context switch across number of process.}}
\label{fig:ctx_switch}
\vspace{-10pt}
\end{figure}

First, the abstraction of secure virtual interpreter (\S\ref{sec:secure_exception_handle}) incurs overhead whenever there is control flow change. Let's consider the case in which \secure process A is stopped (either asynchronously by an interrupt or synchronously by an exception) and \secure process B is resumed on the same CPU, where A and B can be the same (or different threads of a process). The overhead includes the suspension of A and the resumption of B, both of which will invoke the \tcb. When A and B are different, the \ptbar is updated, which traps into the \tcb and incurs further overhead.
We use \program{lat\_ctx} from LMBench to assess these overheads: it measures the latency of context switch among a group of processes. 
\autoref{fig:ctx_switch} shows that \name adds 2.1\textasciitilde2.6~$\mu$s and 3.9\textasciitilde4.7~$\mu$s to non-sensitive and sensitive processes, respectively for each context switch, for 4 to 20 processes.
Because the context switch overhead for non-sensitive processes only include that for \ptbar update, we can estimate it costs 2.1\textasciitilde2.6~$\mu$s in total.
It is nontrivial mostly because \name uses EL2 as the stepping stone to invoke the \tcb (\code{vmcr\_trap()}), which costs about 1.5$\mu$s in a round on our silicon (See \autoref{tab:mode_switch}).
Based on this, we can estimate the overhead in suspension and resumption is about 2~$\mu$s per context switch as is the difference between overheads to sensitive and non-sensitive processes.

Second, the abstraction of secure virtual memory (\S\ref{sec:secure_virtual_mem}) incurs overhead whenever there is any change to any page table in the system and thus the \tcb must get involved. We use \code{lat\_pagefault} from LMBench to assess this overhead with the latency of page fault handling triggered in a mapped file.
The average latency of page fault handling for \code{Vanilla}, \code{Non-sensitive}, and \code{Sensitive} are 1.02~$\mu$s, 1.57~$\mu$s and 1.88~$\mu$s, respectively. 
\name adds 0.55$\mu$s and 0.86$\mu$s to page fault handling for non-sensitive and sensitive processes, respectively.
This overhead includes the cost of trapping into the \tcb and validating the page table update (\S\ref{sec:design-insight}).
The higher overhead for sensitive processes comes from maintaining coherence between user and \shadow page tables (\S\ref{sec:secure_virtual_mem}), and additional validation with the \tcb's bookkeeping about per-process virtual memory areas, in order to defend against memory-mapping related Iago attacks~\cite{checkoway2013iago}.

Third, the overhead from \design can be highly dependent on the silicon, because
it involves three privilege modes, each for user, kernel, and \tcb.
 We devise a microbenchmark to measure the latency for changing CPU privilege modes on our silicon. 
As mentioned in \S\ref{sec:tcb-impl}, \secure process creation and trapping virtual memory control registers both involve EL2 as a stepping stone for invoking the \tcb.
We carefully profile the switching between these modes and find the cost of mode switch can be very different as shown in \autoref{tab:mode_switch}.  Mode switches involving EL2 can be very expensive, which motivates us to implement the \tcb in EL3 and prefer using EL1 as the stepping stone for trapping into EL3 when possible, as explained in \S\ref{sec:tcb-impl}. 

\begin{table}[t!]
\caption{Overhead in \# of cycles for CPU privilege mode change from Row to Column.}\vspace{-10pt}
\label{tab:mode_switch}
\centering
\small
 \begin{tabular}{c||c c c c} 
 Mode & EL0 & EL1 & EL2 & EL3 \\
 \hline\hline
 EL0 & & 44  & 868  & N/A \\ 
 EL1 & 31  & & 1308 & 43  \\
 EL2 & 248 & 16  &  & 817 \\
 EL3 & 17  & 36  & 634  & \\
 \hline
 \end{tabular}
 \vspace{-10pt}
\end{table}

\paragraphc{App performance optimization under \name}:
The overhead analysis above provides important clues into optimizing an application to run as \secure under \name.
First, the high overhead of control-flow change invites techniques that reduce the frequency of control-flow change. System-wise, these include using a lower timer frequency and pinning a \secure thread to a CPU. Because syscalls engender control-flow change through exception, techniques from the literature that reduces syscall exception would also be helpful, e.g., FlexSC~\cite{soares2010flexsc}.
The high overhead of control-flow change also explains the 35.4\% overhead of \secure \program{Redis}. For each request, \program{Redis} makes 6 \code{gettimeofday}, 2 \code{read}, 1 \code{write} and 1 \code{epoll\_wait} syscalls.
So they incur 10 suspension and resumption as well as 16 semantic accesses, adding about 20~$\mu$s to each request.

Second, the high overhead of page fault handling imposes formidable cost to creating a new \secure process: program cold-start engenders a lot of page faults.
This suggests that \design is more efficient at protecting long-running services than ephemeral ones. For example, our original implementations (\program{Original}) of \program{OTP} and \program{DNN} create a new process for each authentication and inference request, respectively. Overhead due to page fault handling (and decrypting segments from binary) contributes to the significantly higher latency for \secure \program{OTP} and \program{DNN}.
Once we re-implement both as daemons (\program{Optimized}), the overhead from \name vanishes as is obvious from \autoref{fig:eval_overhead}.
\subsection{Comparison with Related Systems}

\begin{figure}
\centering
\begin{subfigure}{0.4\textwidth}
\includegraphics[width=\textwidth]{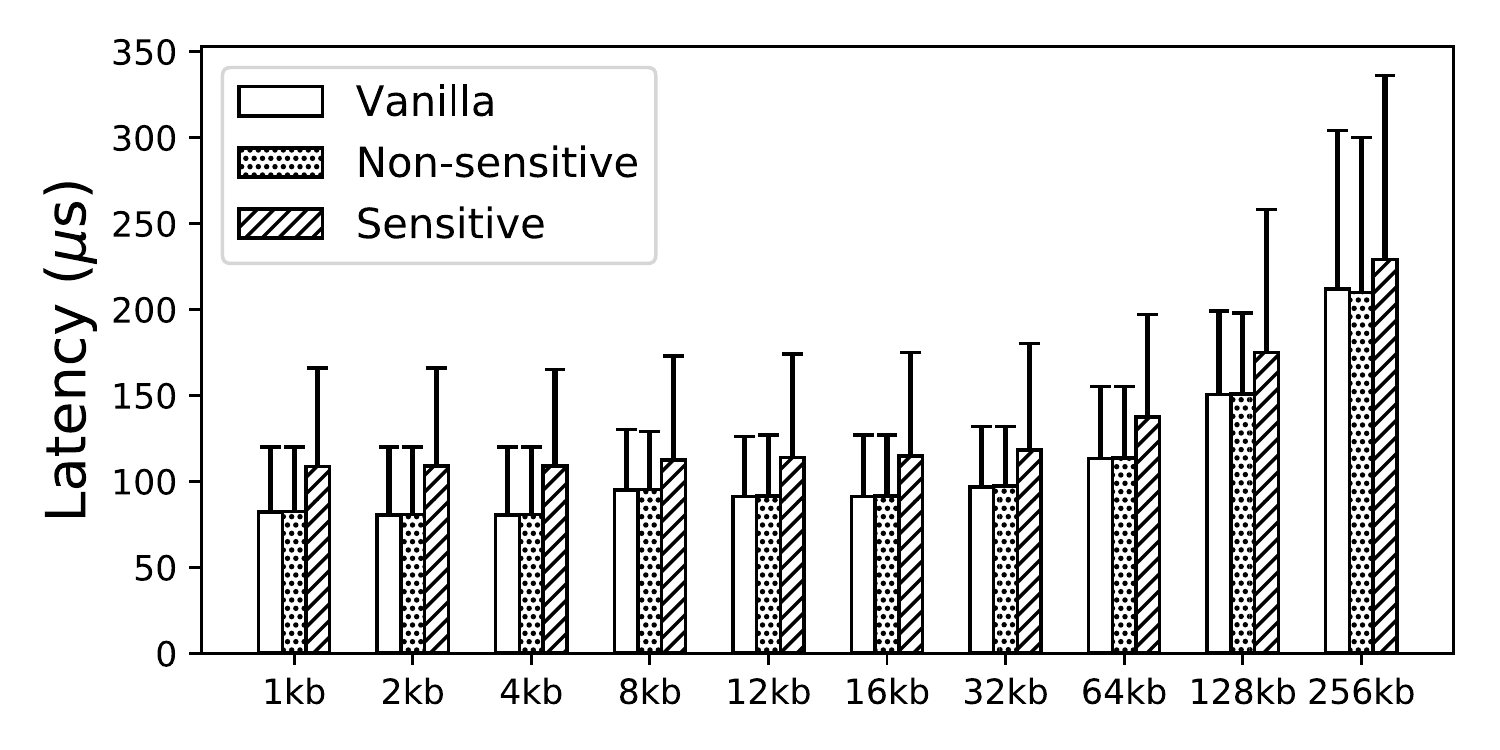}
\vspace{-5ex}
\caption{Local network loop (Y axis linear in $\mu$s)}
\label{fig:eval_nginx}
\end{subfigure}
\begin{subfigure}{0.4\textwidth}
\includegraphics[width=\textwidth]{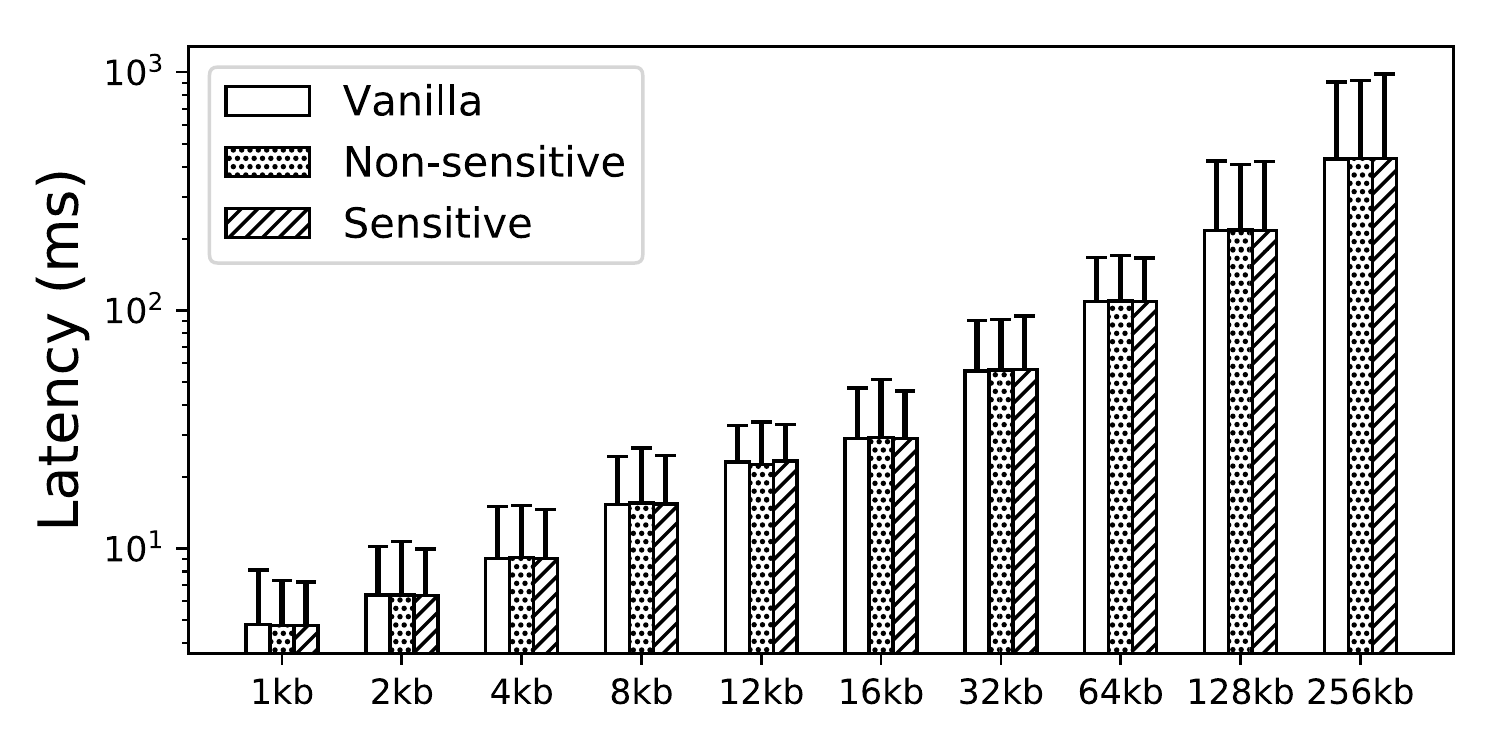}
\vspace{-5ex}
\caption{{Remote client} (Y axis logarithmic in ms)}
\label{fig:eval_nginx_wifi}
\end{subfigure}
\vspace{-2ex}
\caption{\program{Nginx} latency for serving 10 concurrent requests for a HTML file repeatedly in 10 minutes}
\label{fig:nginx}
\vspace{-12pt}
\end{figure}

As discussed in \S\ref{sec:related}, the most related systems are Overshadow~\cite{chen2008overshadow}, TrustShadow~\cite{guan2017trustshadow}, and BlackBox~\cite{van2022blackbox}.
All supports unmodified applications. However, an apple-to-apple performance comparison with any of them turns out to be rather difficult, if possible at all.
In terms of hardware requirement, Overshadow and BlackBox relies on hardware support for nested virtualization; TrustShadow requires ARM-specific hardware feature. In contrast, \name needs neither.
OverShadow supports unmodified OS and unmodified applications but suffers from a massive TCB, i.e., a full-fledged hypervisor. TrustShadow and BlackBox make physical memory used by protected processes disappear from the OS. As a result, they do not support non-semantic access by the OS to a protected user space and OS memory management functions, such as swapping and page migration, cease to support protected applications.

\paragraphb{Deepdive into Overhead with Nginx}
On the other hand, the authors of TrustShadow and BlackBox did report performance overhead for \program{Nginx}, which provides a rare albeit limited opportunity for us to assess \name's performance overhead relatively. 
Based on their designs, we can analyze how \name incurs overhead in a different way from TrustShadow/BlackBox when protecting \program{Nginx}.
\program{Nginx} extensively uses syscalls, i.e., \code{read/write} for network and file operations. 
The overhead of all three systems is proportional to the number of syscalls, because for each syscall, their TCB must get involved as the control flow changes.
The overhead of TrustShadow and BlackBox, however, is also proportional to the size of data exchanged between the user and kernel, because their buffer-based approach require the data to be copied one more time. As a result, the overhead of \name diminishes in percentage as the file size increases, while that of TrustShadow and BlackBox would remain steady. We note the authors of BlackBox only reported the overhead, i.e., 15\%, for a single file size, i.e., 12~KB. The authors of TrustShadow did show the percentage overhead drops as file size increases, however, for a different reason that we will discuss below.

The overhead data reported for TrustShadow and BlackBox were based on a remote client. As a result, their latency measurements include the contribution from the network itself, which can mask much of the protection overhead. This is obvious from comparing the results from \autoref{fig:eval_nginx} and \autoref{fig:eval_nginx_wifi}.
While the overhead of \name's protection is obvious when the client (\program{wrk}) requests files from \program{Nginx} via a local network loop, it varnishes when the client is remote via a high-speed, dedicated Wi-Fi link. (Our board does not have buit-in Ethernet support.) Indeed, the variation from network makes the results statistically indistinguishable for all three cases: \code{Vanilla}, \code{Non-sensitive}, and \code{Sensitive}. 
In contrast, 2\textasciitilde10\% overhead was reported for TrustShadow (1\textasciitilde256 KB files), and 15\% overhead for BlackBox (a 12 KB file).

Finally, the results with remote client also suggest that the overhead from \name's protection may not matter in the grand scheme of things. Similarly, although \name adds several milliseconds of latency to \program{OTP} (\autoref{fig:eval_otp}), it is well below the human-perceptible threshold (several 10s of ms or higher~\cite{myers1985importance,card1991information}). We are hopeful that the overhead of \name (and \design) would be acceptable or barely noticeable in many application scenarios where security of user data matters. 

\paragraphb{Difference in goals and trade-offs}
The major difficulty for an apple-to-apple comparison lies in the difference in our goals. Overshadow, TrustShadow, and BlackBox aim at protecting app data from an untrusted OS. This goal permits them to take short cuts such as making resources used by protected apps disappear from the OS's view (TrustShadow and BlackBox) or relying a huge TCB (Overshadow). \design (and \name), however, aims at rethinking the OS design so that it can manage resources \emph{without} access.

Because of this difference, \design (and \name) must deal with cases that are irrelevant to TrustShadow and BlackBox. 
For example, \design (as well as \name and Overshadow) provides OS an encrypted view into a protected user space so that non-semantic access by the OS such as swapping and page migration still work. 
This encrypted view necessarily incurs the cost of cryptographic operations at runtime.
Such cryptographic operations lead to high overhead, i.e., 25.3~$\mu$s and 44.5~$\mu$s respectively for encrypting/decrypting and signing/verifying a page, in demand paging and any other cases of non-semantic access during runtime. The overhead of demand paging contributes much to the high overhead of process creation as is demonstrated by the original \program{OTP} and \program{DNN}.

Moreover, because of the difference in goals, \design (and \name) must make different tradeoffs, i.e., favoring a smaller TCB (\tcb) and smaller OS modification over performance. For example, without caring about how the OS can be designed differently, 
Overshadow supports legacy OSes with a massive TCB, i.e., a hypervisor.
Both TrustShadow and BlackBox opted for hackish solutions for semantic access that improve performance by allowing a larger TCB and more significant modifications to the OS. 
In contrast, \design employs a single mechanism, i.e., the light weight capability system, to support all semantic accesses. 
This design choice reduces the TCB by 730 SLOC by our estimation as described in \S\ref{sec:tcb-impl}.
In contrast, BlackBox~\cite{van2022blackbox} requires an extra TCB ABI to support \code{futex} and nontrivial OS modifications to support and signal handling.
\name also incurs overhead in a different way: it does not incur extra memory copy or require the TCB to manage memory, but requires a trap into the \tcb for every semantic access. Compared with buffer-based approach used by related systems, \design's design choice results in diminishing percentage overhead as the size of the accessed object grows.
\section{Concluding Remarks}
Modern operating systems (OSes) assume that applications trust them. Hence, they are designed to access any data in user applications. \design shows that the unfettered access to memory is not fundamentally necessary for the OS to perform its own job, including memory management.
\name, a prototype implementation of \design with ARMv8/Linux, leverages a tiny software running at a higher privilege level, called \tcb, to mediate the memory access by the OS. \name shows that modern Unix-like OSes can manage memory without unfettered access or additional virtualization layer. We evaluated its performance with macro and micro benchmarks where we observed that \name provides competitive performance with a runtime TCB 2\textasciitilde3$\times$ smaller compared to previous work.

\section*{Acknowledgments}
\noindent This work is supported in part by NSF Awards \#1730574, \#2130257, and \#2112562.
\newpage
\bibliographystyle{unsrt}
\bibliography{reference/abr-short,reference/reference.bib, reference/security.bib, reference/related.bib, reference/zhong.bib, reference/os.bib}

\end{document}